\documentclass[a4paper,10pt]{article}
\usepackage[utf8]{inputenc}

\usepackage{biblatex} 
\usepackage{amsthm,amsmath,amssymb}

\providecommand{\keywords}[1]
{
  \small
  \textbf{\textit{Keywords---}} #1
}

\author{Luisiana X. Cundin\thanks{Special thanks to: Richard Martindale, Mike Gunter \& Isidore Marshall}}
\title{Newell-Whitehead-Segel Equation: An Exact, Generalized Solution}

\date{September 2, 2024}

\begin{document}

\maketitle

\begin{abstract}
Derivation of an exact, general solution to Newell-Whitehead-Segel transient, nonlinear partial differential equation is provided for one to three dimensional cases, also, arbitrary power of nonlinearity.
\end{abstract}
\section{Introduction}

Nonlinear partial differential equations pose extraordinary difficulty in generating any exact, analytic solutions, due to the intractable nature of the mathematical representations, also, due to the pathological nature of the equations.

The study of nonlinear physical models has existed for well over a century and investigators have developed many approaches to solve such ornery equations by either expansion (perturbation) methods, ingenuous substitutions, reduction methods and, finally, by numerical means or methods; albeit, without knowing an exact solution, the success, accuracy and faithfulness of all such approximation attempts cannot be properly ascertained.

An exact, analytic and most general solution is presented for Newell-Whitehead-Segel transient, one-dimensional equation with arbitrary power of nonlinearity. Specific equations arise for a given integer power of nonlinearity, that is to say, for  a specific nonlinear medium response; thus, several very famous, heavily relied upon nonlinear equations emerge as specific cases of the generalized Newell-Whitehead-Segel equation. These specific equations span many disciplines and are employed in the study of a sundry of physical phenomena, such as, shock waves, nonlinear fluid dynamics; electromagnetic propagation in nonlinear media, nonlinear optics;  \ldots, E\&M transmission through or by cellular media, nerves, neurons in the brain, \ldots, FitzHugh–Nagumo model, Hodgkin-Huxley model; population dynamics in biology, Fisher equation\ldots\&c.

Analysis proves no curvature is possible for Newell-Whitehead-Segel equation, regardless of dimension considered; thus, the only possible solution is the trivial solution, if physical solutions are considered, but purely mathematical investigations may ponder additional results.

\section{Solution}
Consider the generalized Newell-Whitehead-Segel equation, where the power ($n$) of the medium response determines the degree of nonlinearity, \emph{summarily}:
\begin{equation}\label{main}
 \frac{\partial}{\partial t}u-\nu\frac{\partial^2}{\partial x^2}u+\beta u^n=0
\end{equation}

By inspection, the nonlinearity appears in the last term, where the amplitude of the unknown function ($u$) is multiplied by itself $n$ times, that is to say, the amplitude is raised to the power of some positive integer $n$. The domain is considered to be the positive half-plane for time and all space for the spatial dimension.

It is expedient to construct a solution containing Green's function, primarily, due to the majority of the information (energy) contained within the nonlinear partial differential equation accurately represented (captured) by the solution for the corresponding homogeneous, linear partial differential equation, \emph{viz.}:
\begin{equation}\label{sol1}
 u(x,t)=Ge^{-\beta\alpha(t)};G(x,t)=\frac{e^{\frac{-x^2}{4\nu t}}}{\sqrt{4\pi\nu t}},
\end{equation}
\noindent where $\alpha(t)$ is some unknown function of time.

With the bulk of the energy represented by Green's function, the remainder is attributable to nonlinear contributions. To that end, distributing the operators after substituting the assumed solution for the unknown function, $u(x,t)$, yields four terms, where the action of both the first-order temporal derivative and the second-order spatial derivative on Green's function immediately reduces to zero, for Green's function is a known solution to the linear partial differential equation, \emph{viz.}:
\begin{equation}
e^{-\beta\alpha(t)}\left(\frac{\partial}{\partial t}G-\nu\frac{\partial^2}{\partial x^2}G\right)+G\left(\frac{\partial}{\partial t}e^{-\beta\alpha(t)}\right)+\beta \left(Ge^{-\beta\alpha(t)}\right)^n=0
\end{equation}

Since Green's function is a known solution to the resulting equation embraced within the first bracket; cancellation and some simplification yields a first-order, ordinary differential equation in time, where the partial operator can be reduced from a partial to a differential operator, \emph{viz.}:
\begin{equation}
-\beta Ge^{-\beta\alpha(t)}\frac{d\alpha(t)}{dt}=-\beta G^ne^{-n\beta\alpha(t)}
\end{equation}

The resultant $\alpha$-function determines all requirements satisfying the nonlinear term in the original equation, thus, canceling shared terms across the equality, integrating resultant integrals, and isolating for the exponential of the $\alpha$-function yields a general solution for arbitrary degree of nonlinearity, represented by integer power $n$, where the integral of Green's function raised to arbitrary power is left indefinite, but could be transformed into a definite integral, if so desired, \emph{viz.}:
\begin{equation}
 \begin{split}
  -\frac{1}{\beta\left(1-n\right)}\left(e^{-\beta\alpha(t)}\right)^{1-n} & =\int{e^{-\beta\alpha\left(1-n\right)}d\alpha}=\int{G^{n-1}dt};\\
   e^{-\beta\alpha(t)} & =\left(-\beta\left(1-n\right)\int{G^{n-1}dt}\right)^{\frac{1}{1-n}}
 \end{split}
\end{equation}

Upon inspection, the $\alpha$-function has inherited dependence on the spatial variable; therefore, resubstituting the apparent solution, equation (\ref{sol1}), will not cancel all terms generated when operators are applied from the original equation (\ref{main}); in fact, after resubstituting, the operators acting upon the solution in hand, equation (\ref{sol1}), all cancel except the second-order spatial derivative acting upon the $\alpha$-function, which leaves two new, unaccounted for, terms, \emph{viz.}:
 \begin{equation}\label{residual}
 \mathrm{Residue:\ }-2\nu G_x\frac{\partial}{\partial x}e^{-\beta\alpha(t)}-\nu G\frac{\partial^2}{\partial x^2}e^{-\beta\alpha(t)}=0
 \end{equation}

At this point, one may attempt another, more complex solution with additional unknown functions, thereupon, taking another stab at solving all residual terms; but, the process of generating additional residues, no matter the complexity of the assumed solution, will continue \emph{ad infinitum}.

\section{A necessary segue}
Even though the solution obtained, thus far, does not solve the original equation; it is still instructive to analyze the solution, which possesses several key properties. Firstly, A very rich, complex solution has been obtained, where the integral of some arbitrary power of Green's function, in general, is not known; moreover, if the indefinite integral is transformed to a definite integral, the integral yields a family of Exponential integrals, depending on the power of nonlinearity ($n$), furthermore, if the temporal origin is included in the interval of integration, the resultant integral experiences nondeterminate forms in the limit of time to zero. Since the integer value ($n$) is greater than unity, the resulting exponential $\alpha$-function is essentially the $N$-th root of an inverse equation, \emph{viz.}:
\begin{equation}
 \begin{split}
  u(x,t)&=\frac{G}{\left(\beta(n-1)\int{G^{n-1}dt}+C(x)\right)^{\frac{1}{N}}};\\
  &\{N=n-1|n>1\,\&\, n\in\mathbb{Z}^+\},
 \end{split}
\end{equation}
\noindent where a constant of integration $C(x)$ has been explicitly shown.

No restrictions were really made during the derivation of the solution, therefore, the power of nonlinearity could either be a positive integer, fractional number or complex number with real part greater than unity.

Obviously, the linear heat/transport equation results for power of nonlinearity equal to unity, but the general nonlinear solution does not allow for a smooth transition over to linearity. By inspection, the limit of $n$ to unity is a limit representation for the exponential function, if the constant of integration is set to unity, in fact, the limit yields the linear solution, \emph{viz.}:
\begin{equation}
 \lim_{n\rightarrow 1^+}{\frac{1}{\left(\beta(n-1)\int{G^{n-1}dt}+1\right)^{\frac{1}{n-1}}}}\rightarrow e^{-\beta t}
\end{equation}

If the constant of integration, $C(x)$, is greater than unity, the limit yields zero, and, for values greater than zero and less than unity, the limit explodes; as a consequence, the constant of integration should be set equal to unity to maintain rational solutions and a smooth transition from linear to nonlinear medium response models, i.e. physical models. Even though physical constraints force the constant of integration to equal unity; there are no restraints for the constant of integration from a mathematical viewpoint, thus, the solution is very rich and complex, indeed.

Additionally, consider an initial boundary condition involving the impulse function with arbitrary amplitude ($A$), i.e.
\begin{equation}
 u(x,0)=A\delta(x).
\end{equation}

Now, it is known Green's function approaches a Dirac Delta distribution in the limit of time to zero; thus, Green's function contained in the nonlinear solution would too approach a Dirac Delta distribution in the limit of time to zero\textemdash thus, attention shifts to the behavior of the exponential $\alpha$-function in the limit of time to zero. Focusing upon the integrand in the denominator reveals an unknown general anti-derivative with respect to time for Green's function; consequently, the integral will be considered a definite integral with time interval ranging from the origin of time to arbitrary time, $\{t\in[0,t]\}$, further still, the limit of time to origin reduces the interval of integration over the null interval, $[0,0]$, which yields a null result, regardless of the nature of the integrand, \emph{viz.}:
\begin{equation}
 \lim_{\sigma\rightarrow 0^+}\int_0^\sigma{f(t)dt}\rightarrow 0;
\end{equation}

With that said, the definite integral in the denominator of the exponential $\alpha$-function approaches zero in the limit of time to the origin, leaving the constant of integration with arbitrary root set by the power of nonlinearity, \emph{viz.}:
\begin{equation}
 \lim_{t\rightarrow 0}{\frac{BG}{\left(\beta(n-1)\int{G^{n-1}dt}+C(x)\right)^{\frac{1}{n-1}}}}\rightarrow\frac{B\delta(x)}{C(x)^{\frac{1}{n-1}}};
\end{equation}
\noindent where an arbitrary constant $B$ is inserted into the general nonlinear solution.

Finally, we may equate the initial boundary condition to the initial value for the general solution to Newell-Whitehead-Segel equation with the limit of the power of nonlinearity ($n$) to unity, and, by inspection, the only possible value for the constant of integration must be unity, in order to, satisfy the resultant equality, \emph{viz.}:
\begin{equation}
 A\delta(x)=\lim_{n\rightarrow\infty}\frac{B\delta(x)}{C(x)^{\frac{1}{n-1}}}\Rightarrow A\delta(x)
\end{equation}
\noindent where arbitrary constants for amplitudes are equated, i.e. $A=B$.

Even though, for powers of nonlinearity greater than unity, it is certainly possible to set the constant of integration to some arbitrary number in order to equate with the amplitude for the initial boundary condition; but, the spatial dependence is a Dirac Delta distribution and the constant of integration, being a function of the spatial variable itself, is confined to a small point at the spatial origin, hence, it is most convenient to simply set the constant of integration to unity for all powers of nonlinearity.

\section{Another iteration}
The solution generated, so far, adequately removes the nonlinear term involving the unknown function, $u(x,t)$, but fails to properly account for the spatial operator. Cancellation of the nonlinear term has been realized by employing an exponential function of time, which is typically employed for linear versions of the partial differential equation; moreover, the solution generated does, in fact, reduce to the linear solution in the limit of the nonlinear index $n$ to unity, for the residual algebraic terms left upon applying the spatial operator to the solution reduce to zero in the limit of index $n$ to unity, i.e.
\begin{equation}
 \lim_{n\rightarrow 1^+}(n-1)^sR(x,t;n)\rightarrow 0,
\end{equation}
\noindent where index power $s$ is an arbitrary positive integer and general function $R(x,t)$ represents all remaining functions in equation (\ref{res1}) in the limit indicated.

The equation integrated to generate the exponential time function, typically employed to account for the medium response, inheres spatial dependence, in other words, the solution would be exact if the residual equation were a function of time alone. As a consequence, re-introduction of the solution into the partial differential equation cancels most all terms, unfortunately, a set of partial differential terms are generated as a consequence the exponential $\alpha$-function inheritance of spatial dependence, \emph{viz.}:
\begin{equation}\label{res1}
 \begin{split}
  -\nu G_xf_x-\nu Gf_{xx}&=-\nu\beta(n-1) G_xf^n\int{G_xG^{n-2}dt}-\\
  (n-1)^2\beta^2\nu & Gf^{-\frac{2n-1}{n-1}}\left(\int{G^{n-2}G_xdt}\right)^2 -\\
  \nu\beta^2  &\left(n-1\right)^3Gf^{-\frac{2n-1}{n-1}}\left(\int{G^{n-2}G_xdt}\right)^2+\\
  &(n-1)\beta\nu Gf^n\left(\int{(n-2)G^{n-3}(G_x)^2dt}+\int{G^{n-2}G_{xx}dt}\right)
 \end{split}
\end{equation}

The residue or unaccounted information begins to explode in complexity, becoming evermore transcendental in nature. Do notice, as already mentioned, the residue does reduce to a null result in the limit of the nonlinearity index $n$ to unity, which does imply the transition to linearity has been maintained; in other words, the residue is solely attributable to the nonlinearity of the original equation. The residue contains many ornery representations, namely, the derivatives of Green's function, where Green's function solely maintains rationality in the limit of time to the origin, otherwise, any derivative of Green's function becomes unwieldy and swiftly diverges at the time origin.

So, it might be considered fruitful to introduce yet another unknown function into the assumed solution, $u(x,t)=GKe^{-\beta\alpha}$, where the new function ($K$) must also be introduced within the $\alpha$-function, thus,
\begin{equation}\label{nsol}
 u(x,t)=GK(\beta(n-1)\int{G^{n-1}K^{n-1}dt}+1)^\frac{-1}{n-1}
\end{equation}

By introducing a new unknown function, in the manner indicated above, the time derivative of the solution is guaranteed to cancel with the medium response. Unfortunately, the spatial operator would generate additional terms upon acting on all functions, thus, the complexity will swiftly explode. Now, if the newly introduced function $K$ is designated dependent solely on the spatial variable, the resulting equation will be a non-homogeneous, second-order ordinary differential equation with multiple intractable integrals\ldots and, would not be satisfactory, in the least. If time dependence should additionally be imparted to the new unknown function, the resulting non-homogeneous, partial differential equation would too have the nonlinear medium response removed and would offer hope of generating a solution; albeit, the exponential integrals, nature of the algebraic terms and other concerns dash such hopes.

\section{Another stab}
The benefit of the above attempts and analysis, the fundamental nature of the time dependent portion of any solution would involve the fractional representation seen for the $\alpha$-function; therefore, making good use, consider the reduced and simplified solution with simpler spatial function and requisite time dependence that enable removing the nonlinear medium response from the equation, \emph{viz.}:
\begin{equation}
 u(x,t)=\frac{k(x)}{(\beta(n-1)k(x)^{n-1}t+1)^{\frac{1}{n-1}}}
\end{equation}

Once again, taking the derivative with respect to time cancels with the nonlinear medium response, thus, the only remaining term involves the second-order partial derivative with respect to space, \emph{viz.}:
\begin{equation}
 \begin{split}
  -\nu\frac{d^2}{d x^2}\frac{k(x)}{(\beta(n-1)k(x)^{n-1}t+1)^{\frac{1}{n-1}}}&=0\\
  \frac{d}{d x}\frac{k(x)}{(\beta(n-1)k(x)^{n-1}t+1)^{\frac{1}{n-1}}}&=C_1\\
  \frac{k(x)}{(\beta(n-1)k(x)^{n-1}t+1)^{\frac{1}{n-1}}}&=C_1x+C_2
 \end{split}
\end{equation}

Since the partial derivative can be simplified, integration can be realized\textemdash twice over; yielding an identity equating the assumed solution is equal to a linear spatial function. Obviously, the definition will satisfy all requirements for the nonlinear partial differential equation, where the time derivative cancels with the nonlinear medium response (term), and, the second-order spatial derivative is null. Done..

Considering the spatial dependence is linear, therefore, not bounded; all constants of integration must be set equal to zero, leaving the only possible solution to Newell-Whitehead-Segel partial differential equation a trivial solution, $u(x,t)=0$.

\section{Closing}
I, the author, have suspected for some time now\ldots the true solution to nonlinear partial differential equations are in fact null solutions, that is to say, the trivial solution of equaling zero \cite{cundin2020generalsolutiongeneralisednewellwhiteheadsegel}. The reason for this suspicion is based primarily upon Fourier Theory, where convolution is the proper method of multiplying functions together in the original space, consequently, the spectrum of each function should be multiplied together in the codomain. Earlier investigations of nonlinear equations yielded a complex, recursive function involving nested fractions with exponential integrals, thus, making analysis altogether a rather difficult proposition; nonetheless, analysis in the codomian most certainly pointed to either a null result or, in the least, a Dirac Delta distribution in time. This present analysis satisfies my curiosity and reveals simple evidence the solutions to various nonlinear partial differential equations, such as, Burgers' equation, all share one solution in kind, in fact, the trivial solution.

If greater degrees of freedom are suspected to somehow rectify the restrictions imposed by a one-dimensional system\ldots those hopes are dashed, for the solution to Newell-Whitehead-Segel equation in higher dimensions yields similar linear planes as the spatial dependence, \emph{viz.}:
\begin{equation}
 \begin{split}
  \mathbb{R}^2:u(x,y,t)&=\frac{k(x,y)}{(\beta(n-1)k(x,y)^{n-1}t+1)^{\frac{1}{n-1}}}\\
  \mathbb{R}^3:u(x,y,z,t)&=\frac{k(x,y,z)}{(\beta(n-1)k(x,y,z)^{n-1}t+1)^{\frac{1}{n-1}}},
 \end{split}
\end{equation}
\noindent with integrals for each Laplacian operator, respective to dimension, equal to a plane, i.e. $c_1xy+c_2,\text{ or }c_1x+c_2y, \mathbb{R}^2$; for three dimensions, $c_1xyz+c_2\text{ or } c_1x+c_2y+c_3z, \mathbb{R}^3$.

\printbibliography 

\end{document}